\documentclass[12pt,preprint]{aastex}
\slugcomment{To appear in Astrophysical Journal.}
\shorttitle{Collapsed Cores in Globular Clusters}
\shortauthors{Djorgovski et al.}

\begin{document}
\title{New s-process path and its implication to $^{187}$Re-$^{187}$Os
nucleo-cosmochronometer}

\author{T.~Hayakawa\altaffilmark{1,2}, T.~Shizuma\altaffilmark{1}, 
T.~Kajino\altaffilmark{2,3}, S.~Chiba\altaffilmark{4},
N.~Shinohara\altaffilmark{4}, T.~Nakagawa\altaffilmark{4}, T.~Arima\altaffilmark{5}}

\altaffiltext{1}{Advanced Photon Research Center, Japan Atomic Energy Research Institute, Kizu, Kyoto 619-0215, Japan.}
\altaffiltext{2}{National Astronomical Observatory, Mitaka, Tokyo 181-8588, Japan.}
\altaffiltext{3}{Department of Astronomy, Graduate School of Science, University of Tokyo, Bunkyo-ku, Tokyo 113-0033, Japan.}
\altaffiltext{4}{Japan Atomic Energy Research Institute, Tokai, Ibaraki 319-1195, Japan.}
\altaffiltext{5}{Tohoku University, Sendai 980-8578, Japan.}

\begin{abstract}
We study a new s-process path through an isomer of $^{186}$Re 
to improve a $^{187}$Re-$^{187}$Os nucleo-cosmochronometer.
The nucleus $^{187}$Re is produced by this new path 
of $^{185}$Re(n,$\gamma$)$^{186}$Re$^m$(n,$\gamma$)$^{187}$Re.
We measure a ratio of neutron capture cross-sections for the $^{185}$Re(n,$\gamma$)$^{186}$Re$^m$ and
$^{185}$Re(n,$\gamma$)$^{186}$Re$^{gs}$ reactions at thermal neutron energy
because the ratio with the experimental uncertainty has not been reported.
Using an activation method with reactor neutrons, we obtain the ratio of $R_{th}$ = 0.54 $\pm$ 0.11\%.
From this ratio we estimate the ratio of Maxwellian averaged cross sections
in a typical s-process environment at $kT$ = 30 keV with a help of 
the temperature dependence given in a statistical-model calculation
because the energy dependence of the isomer/ground ratio 
is smaller than the absolute neutron capture cross-section.
The ratio at $kT$=30 keV is estimated to be $R_{st}$ = 1.3 $\pm$ 0.8\%.
We calculate the s-process contribution from the new path in a steady-flow model.
The additional abundance of $^{187}$Re through this path is 
estimated to be $N_{s}$ = 0.56 $\pm$ 0.35\% relative to the abundance of $^{186}$Os.
This additional increase of $^{187}$Re does not make any remarkable change in the $^{187}$Re-$^{187}$Os
chronometer for an age estimate of a primitive meteorite, which has recently been found
to be affected strongly by a single supernova r-process episode.
\end{abstract}


\section{Introduction}

Two neutron-capture processes are important for astrophysical nucleosynthesis 
of heavy elements.
The first one is a rapid neutron-capture process (r-process) that is considered
to occur in supernova (SN) explosions \citep{Meyer92,Woosley94,Qi98,Co99,Ot00}, 
and the other is a slow neutron-capture process (s-process) 
in low-mass asymptotic giant branch (AGB) stars \citep{Ka90,Straniero95,Gallino98,Ar99} 
or massive stars \citep{Prantzos90,Woosley95,Th00,Ho01}.
Long-lived radioactive nuclei are used as nucleo-cosmochronometers,
 which are useful for an investigation of 
nucleosynthesis process history along the Galactic chemical evolution (GCE) 
before the solar system formation.
A general idea of the nucleo-cosmochronometer was proposed 
by Rutherford about 70 years ago~\citep{Rutherford}.
Radioactive nuclei of cosmological significance are rare and only six chronometers with half-lives
in the range of the cosmic age 1 $\sim$ 100 Gyr are known.
They are $^{40}$K~\citep{BBFH} and $^{87}$Rb~\citep{Clayton} for the s-process or explosive nucleosynthesis in SNe,
$^{176}$Lu~\citep{Hayakawa2004} for the photodisintegration reaction nucleosynthesis in SNe ($\gamma$-process),
and $^{187}$Re~\citep{Clayton}, $^{232}$Th and $^{238}$U~\citep{BBFH,B2} for the r-process.
Recently two actinoid elements U and Th in a very metal-poor star
were detected for the first time~\citep{Ur}. 
The actinoid nuclei in the old metal-poor star 
were perhaps created in a single r-process of a SN explosion.
Since the initial abundances of $^{238}$U and $^{232}$Th cannot be measured directly
these abundances should be calculated in the framework of the r-process models.
However the calculated results by many models~\citep{Ot00,Goriely01,Schaty02,Wanajo02}
are different from one another and thus
a large uncertainty still remains in the estimation with the U-Th chronometer.

A pair of nuclei, $^{187}$Re-$^{187}$Os, was proposed to be an intentional cosmochronometer to date the 
r-process~\citep{Clayton}.
Fig.~1 shows a nuclear chart 
and flows of nucleosynthesis processes around the nucleus $^{187}$Re.
The nucleus $^{187}$Re is predominantly produced by ${\beta}^{-}$-decay after the freezeout of
the r-process and is located outside the main path of the s-process.
The s-process nuclei $^{186,187}$Os are not directly produced by the r-process
because they are shielded by the stable isobars of $^{186}$W and $^{187}$Re against the $\beta$-decay
after the r-process.
The pure s-process nucleus $^{186}$Os is 
used for the normalization of the s-process abundances in this mass region.
The s-process nucleus $^{187}$Os is also produced by the cosmoradiogenic ${\beta}^-$-decay 
of $^{187}$Re.
The ground state of $^{187}$Re decays to $^{187}$Os 
with a half-life of 4.35$\times$10$^{10}$ yr~\citep{Lindner1986}, 
which is longer than the age of the Universe.
It should be noted that 
the $^{187}$Re-$^{187}$Os chronometer has the advantage 
that applying this chronometer is
free from the uncertainty of the initial abundances calculated by the r-process models.
The epoch from an r-process nucleosynthesis event to the present 
can be evaluated by the present abundances of $^{187}$Re and $^{187}$Os
after the subtraction of the s-process contributions to these nuclei.
Although the initial abundances of $^{187}$Re and $^{187}$Os should be calculated
using the s-process models,
the uncertainty of the calculated s-process abundances 
is generally smaller than those in the r-process models.

The $^{187}$Re-$^{187}$Os chronometer has been applied for studying
the GCE~\citep{Clayton69,Cosner81,Yokoi83,Arnould84}.
This chronometer was also used in analyses of meteorites~\citep{Luck80,Luck83,Birck98}.
The measurements of the $^{187}$Re and $^{187}$Os abundances in iron meteorites
and ordinary chondrites were used for evaluating the ages of the meteorites,
which were formed from a metal reservoir at the solar system formation.
These results lead to an
upper-limit estimation of the age of the Galaxy~\citep{Luck80,Luck83}.
It was pointed out that the half-life of the ${\beta}^-$-decay of  $^{187}$Re
is affected by stellar environments~\citep{Takahashi83}.
The half-life depends on 
the temperature and the electron density of the stellar environments.
The half-life becomes shorter
in high temperature such as $T$ $>$ 10$^8$ K
due to the enhanced bound-state $\beta$-decay in ionized atoms.
This phenomena was verified in
a measurement of the change of the half-life of the ionized $^{186}$Re~\citep{Bosch96}.
However the $^{187}$Re-$^{187}$Os system is still an important chronometer
in the astronomical observations of old stars and in the analyses of pre-solar grains in
primitive meteorites for the chronology of the r-process.

The chemical compositions of the extremely metal-poor stars
are enhanced by the production of heavy elements in neutron-capture nucleosynthesis in the first generation stars.
Recently astronomical observations with isotope separations
have been carried out for several heavy elements such as
Ba~\citep{Lambert02} and Eu~\citep{Sneden02,Aoki03a,Aoki03b}. 
Likewise, the pre-solar grains in the primitive meteorites provide
samples affected strongly by a single nucleosynthesis event.
Heavy elements such as Mo, Te and Ba in the pre-solar grains have
already been separated into isotopes.
The origins of these pre-solar grains are considered to be the ejecta
of core collapse SN explosions~\citep{Richter98,Pellin00,Yin02}.

The purpose of this paper is to present a contribution of a new s-process path
to $^{187}$Re and $^{187}$Os in order to improve the accuracy of this cosmochronometer.
In the previous studies the s-process contributions via two weak paths,
$^{184}$W(n,$\gamma$)$^{185}$W(n,$\gamma$)$^{186}$W(n,$\gamma$) $^{187}$W($\nu$,e$^{-}$)$^{187}$Re
and $^{185}$Re(n,$\gamma$)$^{186}$Re$^{gs}$(n,$\gamma$) $^{187}$Re, were studied.
K{\"a}ppeler {\it et al.} measured the neutron capture cross sections
of $^{185}$Re and $^{187}$Re at $kT$ = 25 keV~\citep{Kappeler1991}.
The neutron capture cross section of an unstable nucleus $^{185}$W
is evaluated from the inverse reaction $^{186}$W($\gamma$, n)$^{185}$W~\citep{Sonnabend,Mohr04}.
Here we propose another new weak s-process branch to produce $^{187}$Re:
a path through a $^{186}$Re isomer which has a half-life of 2.0$\times$10$^5$ yr.
This isomer was found by Lindner about 50 years ago~\citep{Lindner1951},
and then Seegmiller {\it et al.} measured the half-life of this isomer 
by a mass-spectrometric analysis and $\gamma$-ray spectroscopy
of samples after neutron irradiation~\citep{Seegmiller}.
However the absolute value of the neutron-capture cross section for the $^{185}$Re(n,$\gamma$)$^{186}$Re$^{m}$ 
reaction has not been reported.
Seegmiller {\it et al.} reported a ratio of the neutron-capture cross section of $^{185}$Re leading to 
the isomer and the ground state, but the uncertainty was not given in the article.
Therefore the neutron capture cross section to the isomer has not been established~\citep{ToI}.
The half-life of the ground state of $^{186}$Re is only 3.718 days,
while that of the isomer, 2.0$\times$10$^5$ yr, 
is longer than the timescale of the s-process, 10$^{2}$ $\sim$ 10$^{4}$ yr.
The s-process contribution should be, thus, estimated with high accuracy.

In the present work we measure the neutron-capture cross section ratio $^{185}$Re(n,$\gamma$)$^{186}$Re$^{m}$ / 
$^{185}$Re(n,$\gamma$)$^{186}$Re$^{gs}$ at the thermal energy.
An implication of this path in the abundances of 
$^{187}$Re and $^{187}$Os is discussed using our data in a classical steady-flow model,
and an effect to the chronometry is evaluated in a sudden approximation.

\section{Experiment procedure and results}

\subsection{Measurement of neutron capture cross-section ratio}

In order to measure the ratio of the neutron-capture cross-sections 
$^{185}$Re(n,$\gamma$)$^{186}$Re$^{m}$ and $^{185}$Re(n,$\gamma$)$^{186}$Re$^{gs}$
at the thermal energy,
two thin natural Re foils 
were irradiated with thermal neutrons from the nuclear reactor JRR-4 
at the Japan Atomic Energy Research Institute (JAERI).
Two Al-Co wires were also irradiated simultaneously to monitor the neutron flux.
The samples were irradiated by the reactor neutrons for six hours.  
After the irradiation the samples were cooled in a water pool
for about four months in order to reduce the background of the unstable isotopes
with short half-lives.
Four and eight months after the irradiation, $\gamma$-rays emitted from the activated samples
were measured by two HPGe detectors.
In addition sixteen months later, the $\gamma$-rays from one sample 
were measured by one HPGe detector.

The weights of the Re foils are 43.8 mg and 58.1 mg,
whereas the weights of the Al-Co wires are 1.74 mg and 4.58 mg.
The Co element composition is 0.475\%.
The natural Re sample consists of two stable isotopes, namely
$^{185}$Re (37.4\%) and $^{187}$Re (62.6\%), and thus
two unstable isotopes, $^{186}$Re and $^{188}$Re, are mainly produced 
by the neutron capture reactions.
The half-lives of the ground states of these isotopes
are $T_{1/2}=3.718$ days ($^{186}$Re) and $T_{1/2}=17$ hr ($^{188}$Re),
which are much shorter than the $^{186}$Re isomer.
The average flux of the neutrons is estimated to be 4.37$\times$10$^{13}$ cm$^{-2}$s$^{-1}$,
which is evaluated by a measurement of decay $\gamma$-rays
after $\beta$-decay from the activated Co in the wires.
The $\gamma$-rays from $^{28}$Al
cannot be measured
because a cooling time of four months is much longer than the half-life of $^{28}$Al.
The total efficiency of the HPGe detectors is approximately equal to 20\% relative 
to a 3$^{''}{\times}$3$^{''}$ NaI detector
and the energy resolution is 2.2 keV for a 1.3 MeV $\gamma$-ray in $^{60}$Co.
The efficiency is calibrated with standard sources of $^{152}$Eu and $^{133}$Ba.
The $\gamma$-rays from the samples are measured at a 5 cm or 3 cm distance from the 
end caps 
of the detectors in the first or second (third) measurement, respectively,
with a lead-shielded low-background environment.

Fig.~2 shows a partial decay scheme of the $^{186}$Re isomer~\citep{A186}.
The isomer with $J^{\pi}$=(8$^+$) locates at an excitation energy of 149 keV.
This state decays to the ground state through a cascade of $\gamma$-decay.
The ground state subsequently decays to $^{186}$Os or $^{186}$W with $T_{1/2}$=3.718 days.
The $\gamma$-ray transitions from the isomer to the ground state cannot be distinguished
from $\gamma$-rays emitted from other activities.
Strong ${\gamma}$-rays are irradiated from $^{184}$Re ($T_{1/2}$=38 days),
which is produced by a (n,2n) reaction with fast neutrons
existing in the nuclear reactor (see Fig.~3).
Its half-life is comparable to the cooling time,
and hence the intensities are enhanced.
However we observe a 137 keV transition of $^{186}$Os 
(see Fig.~4),
which is the strongest $\gamma$-ray in the decay scheme associated with the isomer.
The yield of the 137-keV $\gamma$-ray is contributed not only from the isomer 
but also from the ground state produced 
directly by the $^{185}$Re(n,$\gamma$)$^{186}$Re$^{gs}$ reaction.

In order to obtain the ratio of the neutron-capture cross section
between the isomer and the ground state, 
we select the cooling time of four months after the irradiation for the first measurement.
After the cooling, the ground state of $^{186}$Re
is still remained because of its large capture cross-section relative to the isomer.
Although the surviving number of the ground state is small, 
the contribution to the 137 keV $\gamma$-ray yield is competitive to the isomer
because the decay rate of the ground state is much higher than the isomer.
In the second and third measurements the ground states are not remained.
The yields of the 137-keV $\gamma$-ray at the second and third measurements
are, thus, proportional to the neutron capture cross section to the isomer.

The number of the nuclei produced by the neutron-capture reaction
is proportional to the $\beta$-decay events, which are
evaluated from the $\gamma$-ray intensity in
the measured spectra by applying the following equation of
%%%
\begin{equation}
N_{total}=\frac{n_{peak}}{R{\cdot}{\epsilon}{\cdot}f_{decay}},
\end{equation}
%%%
\noindent
where $N_{total}$ means the total number of the $\beta$-decays,
$n_{peak}$ denotes the peak counts corresponding to the decay $\gamma$-rays,
$R$ the emission probability of the measured $\gamma$-ray per decay, 
${\epsilon}$ the efficiency
of the HPGe detectors and $f_{decay}$ the correction of the $\beta$-decay during the
measurement time. The $f_{decay}$ is expressed by
%%%
\begin{equation}
f_{decay}={\exp}(-t_1{\cdot}{\lambda})-{\exp}(-t_2{\cdot}{\lambda}),
\end{equation}
%%%
\noindent
where $t_1$ and $t_2$ stand for the start and stop time of the measurement, respectively.
The experimental uncertainty consists mainly of the statistical error 
of the yields of the $\gamma$-rays  and the efficiency of the HPGe detectors.
The $\gamma$-ray energy of 137 keV is near the peak of the efficiency curve of the detectors.
We evaluate the efficiency uncertainty to be approximately 7\% from the deviation
of the $\gamma$-ray efficiency obtained with the calibration sources 
from the efficiency curve estimated by ${\chi}^{2}$ fitting.
The pileup effect of the $\gamma$-rays is negligibly small because the absolute
efficiency is lower than 5\%. The self-absorption of the $\gamma$-rays is also negligibly small.

An energy spectrum of neutrons produced by a nuclear reactor consists of thermal flux 
with energy below 0.2 eV
and epithermal one above.
The rate of the neutron capture reaction is expressed by
%%%
\begin{equation}
N^x=\int n(v) v \sigma(v)_x dv =
{\phi}_{th}{\sigma}_{0}^{x}+{\phi}_{epi}I_0^x,
%={\sigma}_{th} {\phi}_{th}\{1+{\frac{ {\phi}_{epi} }{ {\phi}_{th} }}{\frac{ {\sigma}_{epi} }{ {\sigma}_{th}}}\},
\end{equation}
%%%
\noindent
where $N^x$ denotes the reaction rate for the reaction $x$, 
$n(v)$ the neutron density at the velocity $v$, $\sigma(v)_x$ the cross section of 
the reaction $x$ at the velocity $v$, 
${\phi}_{th}$ and ${\phi}_{epi}$ the thermal and epithermal flux, 
${\sigma}_{0}^{x}$ and $I_0^x$ are the cross section at the thermal energy and 
the resonance integral for the reaction $x$, 
respectively.
A measurement for a sample with a Cd absorber is not carried out in the present experiment
and we cannot obtain the thermal neutron flux. Instead we use an average neutron flux,
which is defined by
%%%
\begin{equation}
{\phi}_{ave}{\equiv}\frac{N_{total}(^{60}\mbox{Co})}
{{\sigma}_{th}(^{60}\mbox{Co})},
\end{equation}
%%%
\noindent
where $\phi_{ave}$ the average neutron flux and 
$\sigma_{eff}^x$ the effective cross section of the reaction $x$ in the reactor neutron spectrum.
The effective neutron-capture cross sections with the average
neutron flux are defined by
%%%
\begin{eqnarray}
{\sigma}_{eff}^{gs}&=&\frac{N^{gs}}{{\phi}_{ave}}
=\frac{ {\phi}_{th}{\sigma}^{gs}_{0}+ {\phi}_{epi} I_0^{gs}}{{\phi}_{ave}}, \\
{\sigma}_{eff}^{m}&=&\frac{N^{m}}{{\phi}_{ave}}
=\frac{ {\phi}_{th}{\sigma}^{m}_{0}+{\phi}_{epi} I_0^{m}}{{\phi}_{ave}}.
\end{eqnarray}
Applying these equations,
we obtain the effective neutron capture cross sections leading to the isomer
of $^{186}$Re as given in Table 1, 
where the mean value derived from the second and third measurements is 0.74 $\pm$ 0.05 barn.
The yield of the $\gamma$-ray at the first measurement consists of 
contributions from the ground-state decay as well as the isomer decay.
The effective neutron capture cross section to the ground state can be evaluated
by subtracting the contribution of the isomer decay, which is calculated from
the average cross section of 0.74 $\pm$ 0.05 barn, from the total yield at the first measurement
(see Table 2).
The obtained cross section to the ground state is 132 $\pm$ 26 barn.
This value is consistent with the thermal values of 118 barn~\citep{Nakajima91} and 
112 barn~\citep{bnl325b} within the uncertainty
and thus the effective cross section derived here is essentially the thermal cross section.
To verify this conjecture, we compare the thermal cross sections and resonance integrals leading to the 
ground state and isomer of $^{186}$Re and their ratio in Table \ref{table:JENDL}.
These values are calculated from Japanese Evaluated Nuclear Data Library (JENDL)
Activation Cross Section File 96~\citep{Nakajima91,JENDL}.
%%%
The ratios, $I_{0}$/${\sigma}_{0}$, for both the ground state and isomer
are approximately equal (see Table 3).
This fact indicates that the resonance integral are written by
\begin{eqnarray}
I_0^{gs} &=& \alpha \sigma_0^{gs},\\
I_0^{m} &=& \alpha \sigma_0^{m},
\end{eqnarray}
with the common coefficient $\alpha$.
The isomer/ground ratio of the capture cross sections at the thermal energy can be derived as
\begin{equation}
\frac{\sigma_0^m}{\sigma_0^{gs}} = 
\frac{\sigma_0^m(\phi_{th} + \alpha \phi_{epi})}{\sigma_0^{gs}(\phi_{th} + \alpha \phi_{epi})}=
\frac{\sigma_0^m \phi_{th} + I_0^m \phi_{epi}}{\sigma_0^{gs} \phi_{th} + I_0^{gs} \phi_{epi}}
= \frac{\sigma_{eff}^m}{\sigma_{eff}^{gs}}, \label{eq:187re_s}
\end{equation}
which verifies our conjecture.
The isomer/ground ratio obtained by applying this equation is $R_{th}$ = 0.54 $\pm$ 0.11\%.
This value is a factor of 1.8 larger than the value of 0.3\% reported in a previous report~\citep{Seegmiller}.
Note that experimental uncertainty was not reported in this report~\citep{Seegmiller}.

There are strong $\gamma$-rays from $^{184}$Re produced
by the $^{185}$Re(n,2n)$^{184}$Re reaction with fast neutrons in Fig.~\ref{fig:largenenough}.
The half-life of $^{184}$Re ($T_{1/2}$=38 days)
is long enough to remain after the cooling time.
This fact suggests the possibility
that the isomer of $^{186}$Re might be also produced by the $^{187}$Re(n,2n)$^{186m}$Re
reaction.
To estimate the contribution by the $^{187}$Re(n,2n)$^{186}$Re$^m$ reaction,
we evaluate the effective cross section of the $^{185}$Re(n,2n)$^{184}$Re reaction
by
%%%
\begin{equation}
{\sigma}_{eff}^{(n,2n)}(^{184}\mbox{Re})=
\frac{ N(^{184}\mbox{Re}) }{ {\phi}_{ave} }.
\end{equation}
%%%
We obtain the result that the effective cross section of the $^{185}$Re(n,2n)$^{184}$Re reaction
is approximately equal to 0.00078 barn.
Since the decay rate of $^{184}$Re is much higher than $^{186}$Re$^{m}$,
the $\gamma$-ray intensities of $^{184}$Re are stronger than 
the $^{186}$Re isomer.
Assuming that the effective cross section of the $^{187}$Re(n,2n)$^{186}$Re
is equal to that of the $^{185}$Re(n,2n)$^{184}$Re reaction,
the contribution to the $^{186}$Re isomer can be calculated by
%%%
\begin{equation}
N(^{186}\mbox{Re}^{m})
=R{\cdot}\frac{ A(^{187}\mbox{Re}) }{ A(^{185}\mbox{Re}) }
{\cdot}{\sigma}_{eff}^{(n,2n)}(^{186}\mbox{Re}){\cdot}{\phi}_{ave},
\end{equation}
%%%
\noindent
where $R$ means the production ratio of the isomer to the ground state of $^{186}$Re,
$A(x)$ means the isotope abundance ratio in the sample and $N(x)$ means the number of the nuclei produced by the neutron-induced reactions.
The contribution of the $^{187}$Re(n,2n)$^{186}$Re$^{m}$ reaction is approximately equal to 0.0007\% relative to
that of the $^{185}$Re(n,$\gamma$)$^{186}$Re$^{m}$ reaction.
We therefore conclude that the contamination due to the $^{187}$Re(n,2n)$^{186}$Re$^m$ reaction is negligibly small.

\section{Discussion}

\subsection{s-Process branch to $^{187}$Re through the $^{186}$Re isomer}

We here estimate the effect of the new path in the s-process 
using a classical steady-flow model~\citep{BBFH,Clayton}.
The steady-flow model was extended by Ward, Newman and Clayton~\citep{Ward76}
to apply to a branching point where the s-process nucleus 
is produced by a neutron capture and/or 
a feeding by $\beta$-decay of a parent nucleus
and also destroyed by a neutron capture and/or $\beta$-decay.
The abundance of the residual nucleus can be calculated by

\begin{equation}
\frac{dN_{j}(A)}{dt}={\sum_{i}}{\lambda}_{i}(A)N^{i}(A)-{\sum_{j}}{\lambda}_{j}(A')N^{j}(A'),
\end{equation}

\noindent
where ${\lambda}$ means the neutron capture rate $\langle$ $n{\cdot}{\sigma \cdot v}$ $\rangle$ 
or the $\beta$-decay rate $\langle$  ln2/T$_{1/2}$ $\rangle$,
$n$ is the neutron number density, T$_{1/2}$ is the half life of the $\beta$-decay,
$N$ means the isotope abundance, 
$A$ and $A'$  are the nuclear mass numbers which have the relationship, $A$=$A'$ (for the $\beta$-decay)
or $A'$-1 (for the neutron capture).
In the limit that the timescale of the s-process is much longer than a mean time of 
neutron capture, $dN/dt$ vanishes and hence
the "steady flow" becomes a good approximation of the s-process.
The equation in this limit is expressed by

\begin{equation}
{\sum_{i}}{\lambda}_{i}(A)N^{i}(A)={\sum_{j}}{\lambda}_{j}(A')N^{j}(A').
\end{equation}

\noindent
Applying this equation to the Re-Os branching point displayed in Fig.~1, we obtain a set of the following equations:
%%%
\begin{equation}
{\lambda}_{{\beta}^-}N_{s}(^{186}\mbox{Re}^{gs})={\lambda}(\mbox{to $^{187}$Os})N_{s}(^{186}\mbox{Os}),\label{eq:12}
\end{equation}
%%%
\begin{equation}
{\lambda}_{{\beta}^-}N_{s}(^{185}\mbox{W})={\lambda}(\mbox{to $^{186}$Re}^{m})N_{s}(^{185}\mbox{Re})+{\lambda}(\mbox{to $^{186}$Re}^{gs})N_{s}(^{185}\mbox{Re}),\label{eq:13}
\end{equation}
%%%
\begin{equation}
{\lambda}(\mbox{to $^{186}$Re}^{gs})N_{s}(^{185}\mbox{Re})=({\lambda}_{{\beta}^-}+{\lambda}_{EC/{\beta}^+})N_{s}(^{186}\mbox{Re}^{gs}),\label{eq:14}
\end{equation}
%%%
\begin{equation}
{\lambda}(\mbox{to $^{186}$Re}^{m})N_{s}(^{185}\mbox{Re})+{\lambda}(\mbox{to $^{187}$W})N_{s}(^{186}\mbox{W})={\lambda}(\mbox{to $^{188}$Re})N_{s}(^{187}\mbox{Re}),\label{eq:15}
\end{equation}
%%%
\begin{equation}
{\lambda}(\mbox{to $^{185}$W})N_{s}(^{184}\mbox{W})={\lambda}_{{\beta}^-}N_{s}(^{185}\mbox{W})+{\lambda}(\mbox{to $^{186}$W})N_{s}(^{185}\mbox{W}),\label{eq:16}
\end{equation}
%%%
\begin{equation}
{\lambda}_{{\beta}^+}N_{s}(^{186}\mbox{Re}^{gs})+{\lambda}(\mbox{to $^{186}$W})N_{s}(^{185}\mbox{W})={\lambda}(\mbox{to $^{187}$W})N_{s}(^{186}\mbox{W}).\label{eq:17}
\end{equation}
%%%
We here assume that the internal transition rate between the ground state and the isomer
is negligibly lower than the neutron capture rate.
The validity of this assumption should be discussed later.
A typical astrophysical nucleosynthesis environment of the s-process is $kT$ = 30 keV and $n$ =  10$^8$ cm$^{-3}$.
Since the following two relations hold, $\lambda$(to $^{188}$Re) $\gg$ ${\lambda}_{{\beta}^{-}}$($^{187}$Re) at $^{187}$Re and
(${\lambda}_{{\beta}^{-}}$ + ${\lambda}_{EC/{\beta}^{+}}$ )($^{186}$Re$^{gs}$) $\gg$ ${\lambda}$(to $^{187}$Re) at $^{186}$Re$^{gs}$, 
on the environmental condition for T$_{1/2}$($^{187}$Re) = 43.5 Gyr and T$_{1/2}$($^{186}$Re$^{gs}$) = 3.718 days,
we omit those two slower reaction rates in Eqs.~(\ref{eq:14}) and (\ref{eq:15}).
The situation does not change when the atom of $^{187}$Re is ionized in the stellar interior during the s-process
and the half-life decreases by 9$\sim$10 orders of magnitude~\citep{Yokoi83,Bosch96}.
Two weak s-process contributions are known~\citep{Kappeler1991}.
The unstable nucleus $^{185}$W is a branching point where it decays to $^{185}$Re with a half-life of 75.1 days
and it is transmuted to $^{186}$W by the neutron capture reaction.
The nucleosynthesis flow from $^{186}$W reaches to $^{187}$Re.
The $\beta$-decay rate at $^{185}$W is higher than the neutron capture rate by an order in the case of $n$=10$^{8}$ cm$^{-3}$,
but it depends on the neutron density in the s-process environment.
The ground state of $^{186}$Re is the second branching point, which decays to both 
$^{186}$Os and $^{186}$W. 
Using Eqs.~(\ref{eq:12})-(\ref{eq:17}), the s-process abundance of $^{187}$Re relative to $^{186}$Os
that is produced by the s-process including the new s-process path is calculated by

\begin{equation}
N_{S}(^{187}Re)=
{\{}
\frac{R}{A}+\frac{ {\lambda}_{{\beta}^+}(^{186}Re^{gs}) }{ {\lambda}_{{\beta}^-}(^{186}Re^{gs}) }
+\frac{ 1+R }{ A }
{\cdot}
\frac{{\lambda}(to ^{186}W)}{ {\lambda}_{{\beta}^-}(^{185}W) }
{\}}
{\cdot}
\frac{ {\lambda}(to ^{187}Os) }{ {\lambda}(to ^{188}Re) }
{\cdot}
N_{S}(^{186}Os),
\label{eq:final}
\end{equation}

\noindent 
where $R$ is defined by ${\lambda}$(to $^{186}$Re$^{m}$)/${\lambda}$(to $^{186}$Re$^{gs}$) and $A$ 
is the branching ratio of the $\beta$-decay of $^{186}$Re$^{gs}$, 
namely $A$ = ${\lambda}_{{\beta}^-}$/(${\lambda}_{{\beta}^-}$+${\lambda}_{EC/{\beta}^+}$).
We use the branching ratio of 0.9253 for $A$~\citep{A186}.
It should be noted that 
the s-process abundance of $^{187}$Re
has two terms proportional to the ratio of the neutron capture rates leading to the isomer and the ground state of $^{186}$Re.

The Maxwellian-average neutron-capture cross section (MACS) with quasi-stellar neutrons 
corresponding to an energy range of $kT$=8$\sim$30 keV
is important for calculating nucleosynthesis under typical s-process environments.
We measure only the neutron-capture cross-sections with the thermal neutrons in the present experiment.
In order to estimate the s-process abundance of $^{187}$Re,
we assume first that the ratio of neutron capture rates,
$\lambda$(to $^{186}$Re$^m$)/$\lambda$(to $^{186}$Re$^{gs}$) = 0.54 $\pm$ 0.11\%, does not
depend on the neutron energy.
The MACS at $kT$=30 keV for Re and Os isotopes
are taken from the best set evaluated previously~\citep{Kappeler1991}.
The values are 1160 and 418 mbarn for $^{187}$Re and $^{186}$Os, respectively.
We assume the neutron density of 10$^{8}$ cm$^{-3}$ 
to estimate a neutron capture rate of ${\lambda}$(to $^{186}$W).
Substituting the measured cross section ratio, $R_{th}$ = 0.54\%, into Eq.~\ref{eq:final},
we obtain the s-process abundance of $^{187}$Re of $N_{s}$ = 0.23 $\pm$ 0.05\% relative to $^{186}$Os.

As an alternative method to estimate the isomer/ground ratio at stellar energy, 
we calculate the energy dependence of the ratio by the Hauser-Feshbach statistical-model,
which was used in JENDL Activation Cross Section File~\citep{Nakajima91,JENDL}.
In JENDL the total capture cross section
is calculated using a resonance formula and the statistical-model (see Fig.~5).  
We first calculate the capture cross section to the ground state,
which is normalized at 100 keV to the value given in JENDL.
The isomer/ground ratio is also given in JENDL, but its value at the thermal
energy is smaller than the presently measured value by a factor of 23.  
In order to reproduce the measured isomer/ground ratio at the thermal energy,
we adjust the parameters in the present statistical-model calculation in two different ways.
In the first method we adjust the $E3$
transition probability in the $\gamma$-decay cascade to enhance the production
of the isomer.  In the second method we adjust the level density parameters (LDP)
which affects the branching of the neutron and $\gamma$ emission 
and affects the average number of cascade $\gamma$-transitions.
The calculated isomer/ground ratios are shown in Fig.~6 as a function of the incident neutron energy.  
The fact that the energy dependence of the two calculated results is almost identical 
indicates that the calculated ratio is not sensitive to the choice of the parameters.  
Finally both results give the isomer/ground ratio of the MACS at $kT$=30 keV to be $R_{st}$ = 1.3 $\pm$ 0.8\%.  
Although the two calculated ratios are identical, 
we assign an error of 50\% to the calculated result
as a conservative estimate of the uncertainty.   
By inserting this ratio, $R_{st}$ = 1.3 $\pm$ 0.8\%, to Eq.~\ref{eq:final}
we obtain the s-process abundance of $^{187}$Re, $N_{s}$ = 0.56 $\pm$ 0.35\%, relative 
to $^{186}$Os.
This value is larger than that at the thermal energy by a factor of 2.4.
This difference shows that the energy dependence of the ratio is important for the
study of the s-process
and an experimental measurement of the cross section at the stellar energy
is a further subject.

Fig. 6 shows that the calculated ratio increases drastically above 10 keV.  
This tendency indicates that the abundance ratios for the Re and Os isotopes 
can be used as a nucleo-thermometer of the s-process
due to the strong temperature dependence of the isomer/ground ratio.

Finally we would like to discuss a problem concerning
the transition probability between the isomer
and the ground state. 
We assume that this transition probability on a typical s-process condition
is ignored.
A case of the nucleus $^{180}$Ta gives a clue for this problem.
The nucleus $^{180}$Ta has a similar nuclear structure: 
the isomer is meta-stable with $J^{\pi}$=9$^+$ ($T_{1/2}$ $>$ 10$^{15}$ yr) 
existing in the solar system,
while the ground state with $J^{\pi}$=1$^+$ 
is unstable against the $\beta$-decay with a half-life of 8.15 hr.
This nucleus is proposed to be synthesized through a weak branch of the 
s-process~\citep{Yokoi83b,Belic99}
and by photodisintegration reactions in supernova explosions~\citep{Around03,Utsunomiya03}.
The isomer synthesized by either process may decay to the ground state
through intermediate states (IMS)
located above the isomer 
by ($\gamma$,${\gamma}'$) reactions
in high temperature environment,
and subsequently the ground state decays to daughter nuclei through $\beta$-decay.
The residual abundance of $^{180}$Ta is affected by the transition probability between the ground state and the isomer,
which depends on the excitation energy of the IMS~\citep{Belic99}.
It is pointed out that the $^{180}$Ta abundance contributed by the s-process should be smaller
than the present solar abundance
if the IMS might exist below excitation energy of 1 MeV~\citep{Th00}. 
This suggests that there is no IMS lower than 1 MeV.
The IMS have been investigated experimentally,
but the IMS below 1 MeV has not been found.
Recently the transition probability from the isomer to the ground state
has been directly measured in photo-excitation experiment~\citep{Belic99}
and the result supports the previous investigation.

In the case of $^{186}$Re, the IMS at low excitation energy also play an essential role for 
the destruction of the synthesized $^{186}$Re isomer,
but these IMS of $^{186}$Re have not been found.
The transition probability between the ground state and isomer via IMS
depends on the quantum number $K$~\citep{Beer81}.
The quantum number $K$ is a projected component of the total angular
momentum on the nuclear symmetry axis.
The transition probability increases with increasing the deformation of the
nuclear shape~\citep{Carroll91}, 
because the Coriolis interaction in high spin states of the deformed nuclei
causes the mixing of the states with the different $K$ values.
Since $^{180}$Ta is more distant from the magic number,
the deformation of $^{180}$Ta is larger than $^{186}$Re~\citep{shape90}.
This fact indicates 
that the effect of the Coriolis interaction on $^{186}$Re is smaller than that of $^{180}$Ta.
We therefore assume that the transition between the ground state and the isomer via the IMS 
can be ignored in the typical s-process environment.
The measurement of the transition probability of $^{186}$Re may be important 
for the study of the s-process abundance of $^{187}$Re.
The direct measurement of the transition probability is difficult
because the isomer of $^{186}$Re does not exist in the solar system.
A search for excited states by in-beam $\gamma$-ray spectroscopy
such as a measurement of excited states of $^{187}$Re~\citep{Shizuma03}
would be a further subject.

\subsection{Effect on $^{187}$Re-$^{187}$Os chronometry}

The $^{187}$Re - $^{187}$Os chronometer is useful for the study of the age estimate 
of an r-process event.
The purpose of this subsection is to discuss the effect of the new s-process path 
on the $^{187}$Re - $^{187}$Os chronology.
The contribution from the new s-process path through the $^{186}$Re isomer has been totally 
ignored in the past.
We first present the age estimate  without the effects of all weak s-process branchings.
In such a case an age of an object which exhibits both the $^{187}$Re and $^{187}$Os abundances is calculated by

\[
T =
-\frac{T_{1/2}(^{187}\mbox{Re})}{\ln2}{\times}
\]
\begin{equation}
\ln\Biggl(\frac{N_{ob}(^{187}\mbox{Re
})}
{ N_{ob}(^{187}\mbox{Os}) - N_{S}(^{187}\mbox{Os}) + N_{ob}(^{187}\mbox{Re}) }\
\Biggr),\label{eq:age1}
\end{equation}

\noindent
where $N_{ob}(A)$ means the observed abundance and $N_{S}(A)$ means the abundance contributed from the s-process. 
Note that we here adopt a simple model taking a sudden approximation that the r-process occurred at 
the look back time at $t$ = $T$ only once.
Although the sudden approximation is not generally a good approximation for analyses of the solar materials,
it is still useful in analyses of primitive meteorites that have recently been found to be affected strongly 
by a single supernova r-process episode~\citep{Amari92,Yin02}. This approximation also applies to an analysis of
astronomical date such as the abundances of the r-process elements detected
in metal-poor stars with isotope separation
because the elements on the surface of the old low-metallicity stars are probably generated 
in a single nucleosynthesis event~\citep{Sneden96,Ur,Otsuki03,Honda04}.

We present the age estimation by taking into account the new s-process path to $^{187}$Re. In this case,
Eq.~(\ref{eq:age1}) should be replaced by

\[
T =
-\frac{T_{1/2}(^{187}\mbox{Re})}{ln2}{\times}
\]
\begin{equation}
\ln\Biggl(\frac{ N_{ob}(^{187}\mbox{Re}) - N_{S}(^{187}\mbox{Re}){\cdot}{\exp}(-T_{S}{\cdot}{\ln}2/T_{1/2}(^{187}Re) ) } 
{ N_{ob}(^{187}\mbox{Os}) -  N_{S}(^{187}\mbox{Os}   ) + (N_{ob}(^{187}\mbox{Re}) - N_{S}(^{187}\mbox{Re}) ) }\
\Biggr),\label{eq:age2}
\end{equation}

\noindent
where we assume that the s-process occurred at different look back time $T_{S}$ from the r-process episode, i.e. $T{\ne}T_{S}$,
$N_{S}(A)$ means the contribution by the s-process, $N_{ob}(A)$ means the observed abundance.
The s-process contribution to $^{187}$Os should be modified
by the change of the nucleosynthesis flow at $^{186}$Re.
The s-process contribution to $^{187}$Re presented in Eq.~(\ref{eq:final}) 
is calculated from Eqs.~(\ref{eq:12}) $\sim$ (\ref{eq:17}), which contain all the weak s-process paths.
In Eq.~(\ref{eq:final}) the contribution of the new s-process path is expressed by the term of $R$.

Substituting the estimated MACS ratio of $R_{st}$ = 1.3\%  at $kT$= 30 keV,
we obtain the results shown in Tables \ref{table:age1} and \ref{table:age2}.
The neutron capture cross sections at $kT$ = 30 keV are taken from the previous article~\citep{Kappeler1991}.
Those of $^{186,187}$Os and $^{187}$Re are 418, 874, 1160 mbarn, respectively.
We use the solar abundances for $N_{ob}$($^{187}$Re) and $N_{ob}$($^{187}$Os).
$N_{S}$($^{187}$Os) is calculated from the observed abundance $N_{ob}$($^{186}$Os) = $N_{S}$($^{186}$Os).
The $N_{ob}$($^{186}$Os)/$N_{ob}$($^{187}$Re)
ratios in the primitive meteorites or the metal-poor stars 
may be different from the solar abundance ratio
because $N_{ob}$($^{187}$Re) should be enhanced
in the sample which is strongly affected by the single r-process event.
We thus introduce a parameter, $f$, which is defined by
\begin{equation}
N_{ob}(^{186}\mbox{Os})/N_{ob}(^{187}\mbox{Re})=f{\times}N_{\odot}(^{186}\mbox{Os})/N_{\odot}(^{187}\mbox{Re}),\label{eq:define}
\end{equation}
for allowing the variation of $N_{ob}$($^{186}$Os)/$N_{ob}$($^{187}$Re).

The difference between the calculated ages for $R$ = 0 and 1.3\% is
at most $\sim$ 1\% in either case $T_{S}$=5$\times$10$^{9}$ yr (Table \ref{table:age1})
or 10$^{10}$ yr (Table \ref{table:age2}).
The ages are slightly sensitive to the neutron density.
Aoki et al. reported isotope ratios of Eu in s-process element rich metal-poor stars
and show the possible neutron density range of the s-process environment
by comparing their data with the s-process calculation: 10$^{7}$ $\le$ n $\le$ 10$^{9}$ cm$^{-3}$
\citep{Aoki03a,Aoki03b}.
Thus we also present the age for the neutron density of 10$^{7}$ cm$^{-3}$ as well as 10$^{8}$
cm$^{-3}$ in Tables \ref{table:age1} and \ref{table:age2}.
The difference between the ages for two different neutron densities is at most $\sim$ 2\%.
Several metal-poor halo stars are thought to form from either the s-process element rich gas \citep{Aoki03a}
or the r-process element rich material \citep{Sneden96,Ur,Honda04}, which are ejected from 
progenitor stars. The parameter $f$ can be even larger or smaller than unity.
The primitive meteorites, which are affected by a single supernova r-process,
can have the $f$ values smaller than unity.
The column of $f$ in Tables \ref{table:age1} and \ref{table:age2} shows this parameter.
The estimated ages for $R$ = 0 and 1.3\% under the conditions of $f <$ 1.0
show almost the same results because the absolute nucleosynthesis flow from the s-process decreases with decreasing 
the $f$ value. 
In particular, the ages become slightly different with increasing $f$ value for the case of $T_{S}$=10$^{10}$ yr (Table \ref{table:age2}).
Although the effect of the new s-process path seems to be small,
the effect is almost the same as those of the other s-process paths
under the condition of $f$ $\le$ 1.0.
Therefore the direct measurement of the absolute neutron capture cross section of $^{185}$Re leading to the isomer of $^{186}$Re 
at $kT$=8$\sim$30 keV is important for the $^{187}$Re - $^{187}$Os chronometer.

\section{Summary}
%%%
We propose a new s-process path that leads to synthesize $^{187}$Re and 
$^{187}$Os, which form an important nuclear-pair as a nucleo-cosmochronometer of the r-process.  
The new path consists of a neutron capture reaction chain from $^{185}$Re to the isomer of $^{186}$Re
and from the $^{186}$Re isomer to $^{187}$Re, namely
$^{185}$Re(n,$\gamma$)$^{186}$Re$^m$(n,$\gamma$)$^{187}$Re, which has been 
ignored in all previous studies.  
We measure the ratio of the reaction cross sections 
$^{185}$Re(n,$\gamma$)$^{186}$Re$^m$ and $^{185}$Re(n,$\gamma$)$^{186}$Re$^{gs}$
with thermal neutrons provided by a nuclear reactor at the JAERI.
We measure $\gamma$-rays after the $\beta$-decay from the Re samples by HPGe detectors
and thereby we obtain the ratio of $R_{th}$ = 0.54 $\pm$ 0.11\%.
The Maxwellian averaged cross section reaction ratio at the stellar energy 
$kT$ = 30 keV is estimated with the help of the statistical model calculation.  
The result is used to 
calculate the s-process abundance of $^{187}$Re relative to the abundance of $^{186}$Os in the 
classical steady-state flow model, which yielded a value of $N_{s}$ = 0.56 $\pm$ 0.35\%.  
The proposed new s-process path does not make any remarkable change in the $^{187}$Re-$^{187}$Os chronometer
in the sudden approximation when all other possible s-process branchings are included in the calculation.
This confirms the robustness of the $^{187}$Re-$^{187}$Os chronometer which applies to the age estimated
by the analyses of primitive meteorites that are presumed to be affected strongly
by a single supernova r-process episode.
Since the $^{187}$Re atom in these meteorites most likely has not been astrated in the stellar interior
after it was produced in the r-process,
the chronometer is free from the change of the half-life of ionized $^{187}$Re.
We however need further experimental study to measure the cross section
precisely at the stellar energy $kT$=8$\sim$30 keV.  

\begin{acknowledgments}

We would like to thank K. Takahashi, H. Utsunomiya and M. Fujiwara for valuable discussions.
We also thank the crew of the nuclear reactor at JAERI.
This work has been supported in part by Grants-in-Aid for Scientific
Research (15740168) of 
the Ministry of Education, Culture, Sports, Science and Technology of Japan, 
and the Mitsubishi Foundation.

\end{acknowledgments}

%%%Facilities: \facility{Nickel}, \facility{HST(STIS)}, \facility{CXO(ASIS)}.

\clearpage

\begin{table}
\caption{Effective neutron-capture cross sections of $^{185}$Re 
leading to the $^{186}$Re isomer.
Two samples were irradiated for about six hours by reactor neutrons. Time means the cooling time
after the irradiation. Since the cooling time is long enough, the ground state with a short half-life
does not survive. }
\begin{tabular}{lllll} 
\hline
Sample & Time          &  $\sigma_{eff}^{m} $ (barn) & error (barn) \\
       & (month)       &                     & \\
\hline
1  & 8      & 0.78 & 0.06 \\
& 12      & 0.60 & 0.08 \\
2   & 8     & 0.84 & 0.10 \\
\hline
Average & & 0.74 & 0.05\\
\hline
\end{tabular}
\label{table:isomer}
\end{table}

\begin{table}
\caption{Effective neutron capture cross section of $^{185}$Re 
leading to the ground state of $^{186}$Re.
Two samples
were irradiated for about 6 hours by reactor neutrons. Time means the cooling time
after the irradiation. The capture cross section was obtained by subtracting the 
contribution of the isomer decay. }
\begin{tabular}{lllll} 
\hline
Sample & Time          &  ${\sigma}^{gs}_{eff}$ (barn) & error (barn)\\
       & (month)       &                     & \\
\hline
1  & 4      & 136 & 30 \\
2  & 4      & 142 & 52 \\
\hline
Average & & 132 & 26 \\
\hline
\end{tabular}
\label{table:ground}
\end{table}

\begin{table}
\caption{Neutron capture cross-sections at thermal energy, the resonance integrals and their ratios
for the $^{185}$Re(n,$\gamma$)$^{186}$Re$^{gs}$ and $^{185}$Re(n,$\gamma$)$^{186}$Re$^{m}$ reactions calculated
from JENDL Activation Cross Section File 96}
\begin{tabular}{lllll} \hline
            &  $\sigma_0$ (barn)  & $I_0$ (barn) & $I_0/\sigma_0$ \\ \hline
gs        & 117.6          & 1850  & 15.7 \\
isomer        & 0.02711          & 0.443 & 16.3 \\
\hline
\end{tabular}
\label{table:JENDL}
\end{table}

\begin{table}
\caption{Comparison of the ages in units of Gyr calculated using the $^{187}$Re - $^{187}$Os chronometer
with (R = 1.3\%) and without (R = 0\%) the new s-process path.
The age of the s-process, $T_{S}$=5$\times$10$^{9}$ yr, is assumed.
$f$ is a parameter to allow the variation of observed $^{186}$Os/$^{187}$Re ratio relative to the solar ratio (see Eq.~(\ref{eq:define}) )
and n is the neutron density in units of cm$^{-3}$.
R means the ratio of the neutron capture
leading the isomer and the ground state of $^{186}$Re.}
\begin{tabular}{lllll} \hline
$f$         &  Eq.~(\ref{eq:age1})  & Eq.~(\ref{eq:age2}), n=10$^{8}$, R=0\% & Eq.~(\ref{eq:age2}), n=10$^{8}$, R=1.3\% &   Eq.~(\ref{eq:age2}), n=10$^{7}$, R=1.3\% \\ 
\hline
1.5        & 4.86       & 4.86    & 4.86  & 4.86 \\
1.2        & 7.73       & 7.80    & 7.80  & 7.76 \\
1.0        & 9.56       & 9.66    & 9.67  & 9.61 \\
0.5        & 13.9       & 14.0    & 14.1  & 14.0 \\
0.1        & 17.2       & 17.3    & 17.3  & 17.2 \\
\hline
\end{tabular}
\label{table:age1}
\end{table}

\noindent
\begin{table}
\caption{The same as those in Table \ref{table:age1}, but for $T_{S}$=10$^{10}$ yr}
\begin{tabular}{lllll} \hline
$f$         &  Eq.~(\ref{eq:age1})  & Eq.~(\ref{eq:age2}), n=10$^{8}$, R=0\% & Eq.~(\ref{eq:age2}), n=10$^{8}$, R=1.3\% &   Eq.~(\ref{eq:age2}), n=10$^{7}$, R=1.3\% \\ 
\hline
1.5        & 4.86      & 4.69   & 4.68      & 4.79 \\
1.2        & 7.73      & 7.67   & 7.66      & 7.70 \\
1.0        & 9.56      & 9.55   & 9.55      & 9.56 \\
0.5        & 13.9      & 14.0   & 14.0      & 14.0 \\
0.1        & 17.2      & 17.2   & 17.2      & 17.2 \\
\hline
\end{tabular}
\label{table:age2}
\end{table}

\begin{figure}
{\includegraphics[viewport=10mm 75mm 170mm 190mm, clip, scale=1.0]{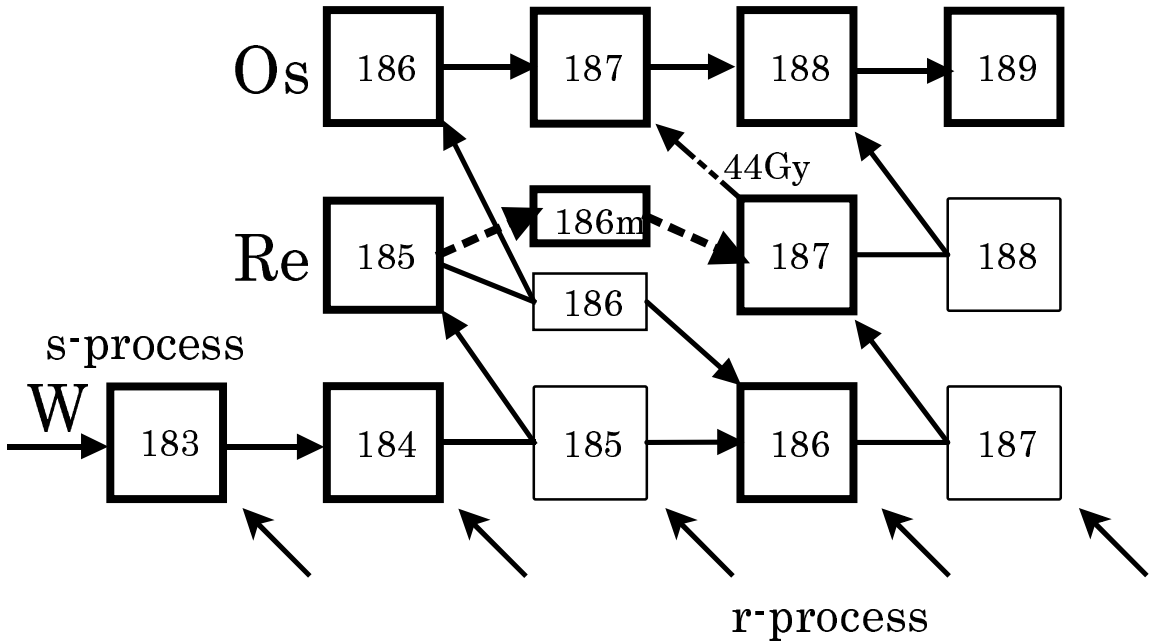}}
\caption{Nuclear chart and nucleosynthesis flow around $^{187}$Re. 
The nuclei in this mass region are synthesized by the s- and r-processes.
The ground state of $^{187}$Re decays to $^{187}$Os with a half-life of about 44 Gyr.
The dotted-line is the new s-process path to $^{187}$Re proposed in this work. }
\label{fig:chart}
\end{figure}

\begin{figure}
{\includegraphics[viewport=10mm 90mm 160mm 170mm, clip, scale=1.0]{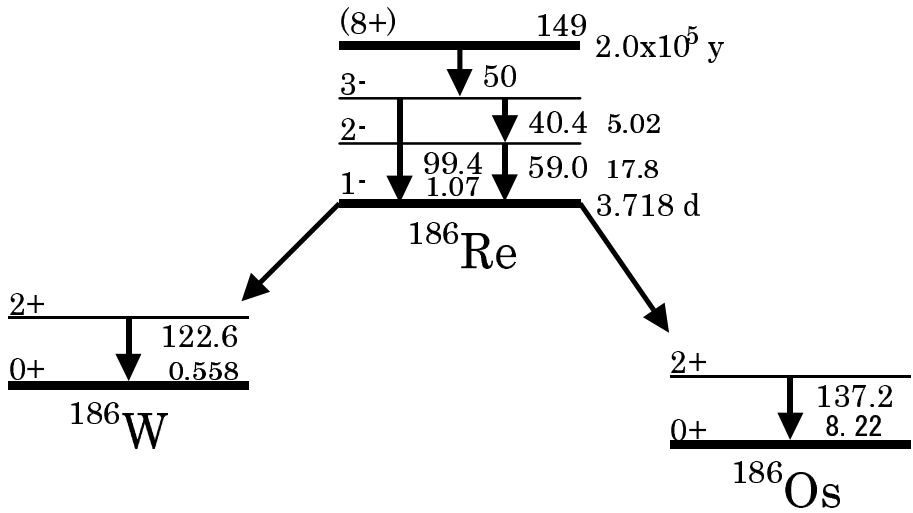}}
\caption{Partial decay scheme of the $^{186}$Re isomer. Relative ${\gamma}$-ray intensities
are indicated. Large number means the energy of the $\gamma$-ray
and small number means the relative intensity.
The isomer of $^{186}$Re decays to the ground state through the internal transitions
and subsequently the ground state disintegrates to $^{186}$Os or $^{186}$W.}
\label{fig:level}
\end{figure}

\begin{figure}
{\includegraphics[viewport=0mm 102mm 220mm 220mm, clip, scale=0.8]{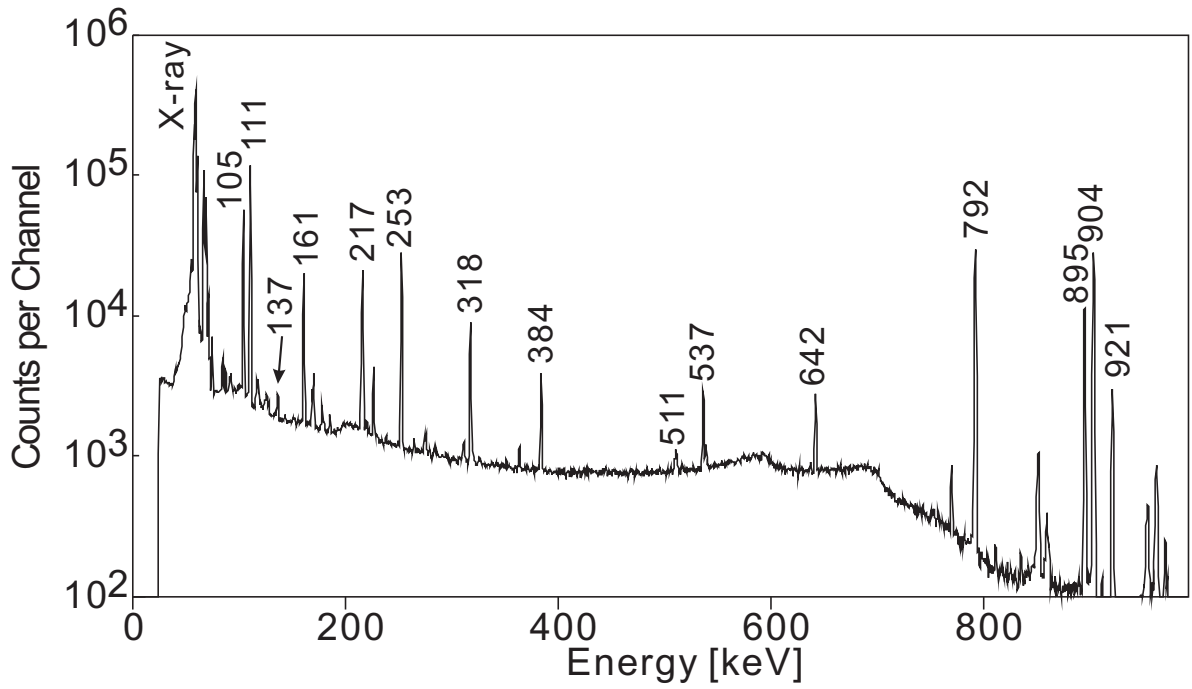}}
\caption{Spectrum of $\gamma$-rays from the $^{nat}$Re target. Strong $\gamma$-rays from
$^{184}$Re are observed, which is produced by the $^{185}$Re(n,2n)$^{184}$Re reaction.}
\label{fig:largenenough}
\end{figure}

\begin{figure}
{\includegraphics[viewport=5mm 105mm 210mm 210mm, clip, scale=0.8]{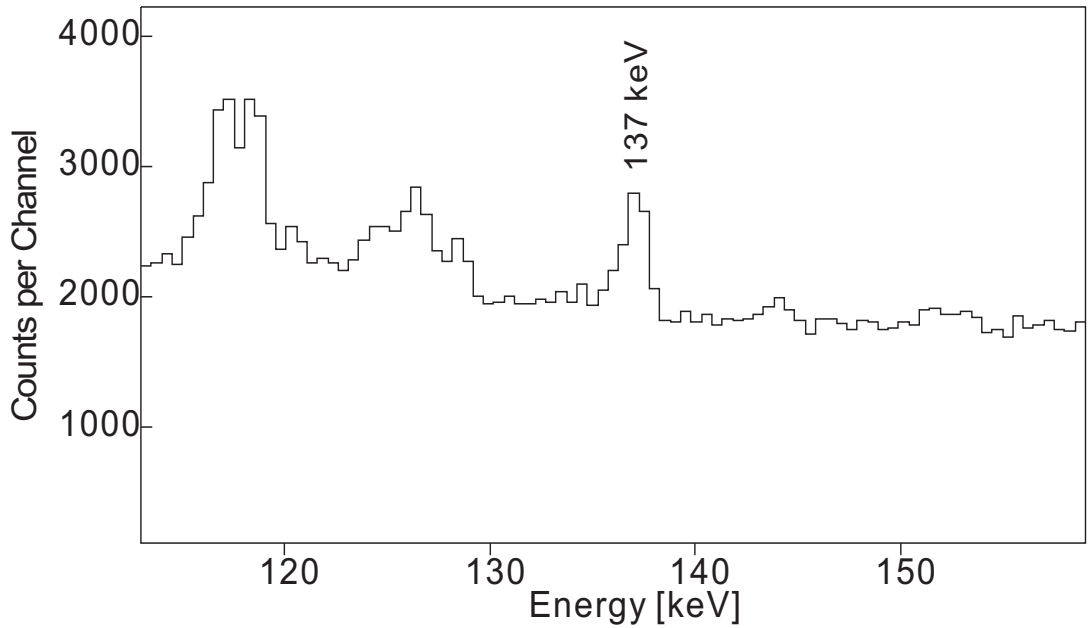}}
\caption{The $\gamma$-ray of $^{186}$Os. The 137 keV $\gamma$-ray
is clearly observed.}
\label{fig:toosmall}
\end{figure}

\begin{figure}
{\includegraphics[viewport=40mm 10mm 260mm 230mm, clip, scale=0.8]{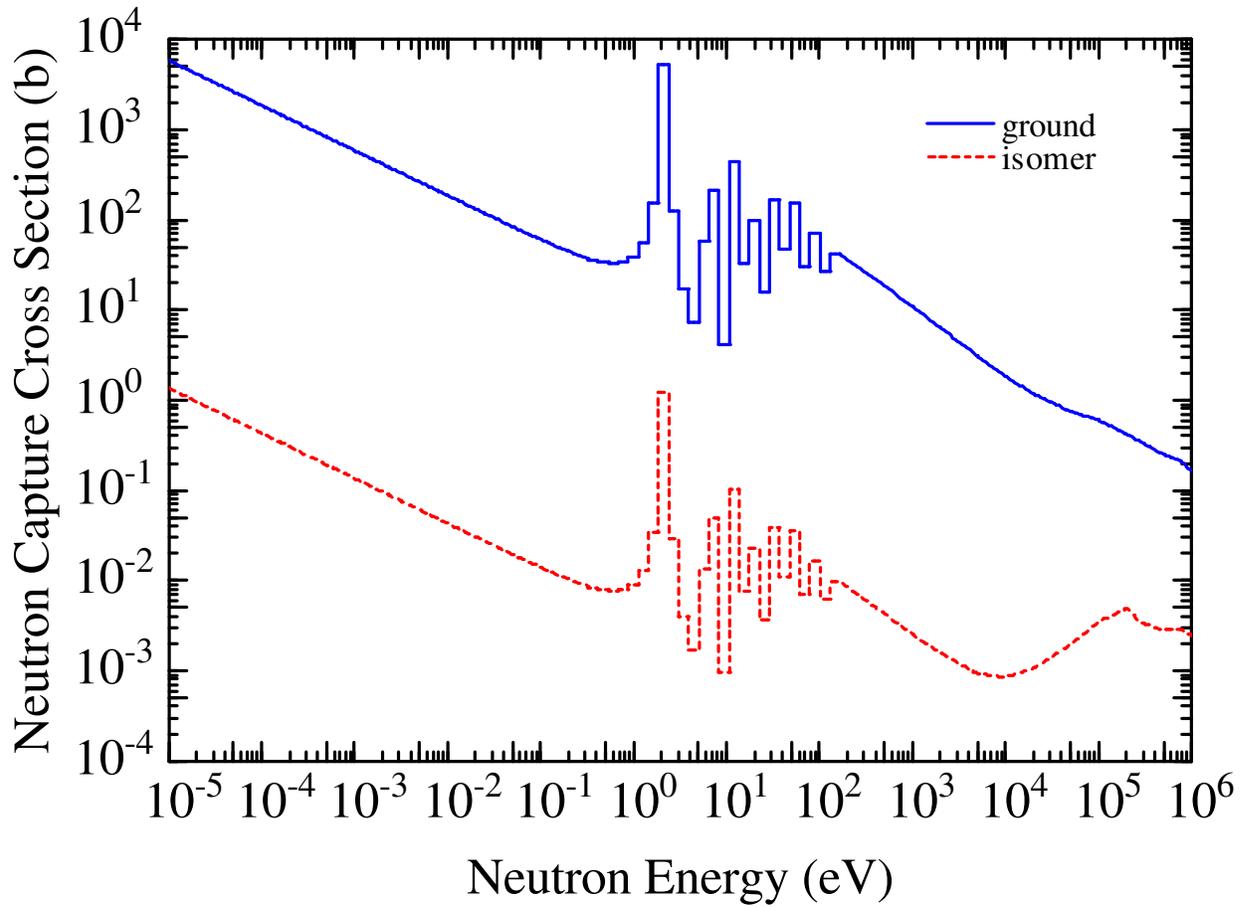}}
\caption{Evaluated neutron capture cross section data for the $^{185}$Re(n,$\gamma$) reaction
leading to the ground state (solid line) and isomer (dashed line) of $^{186}$Re given in 
JENDL Activation Cross Section File 96.}
\label{fig:jendl}
\end{figure}

\begin{figure}
%%%%{\includegraphics[viewport=0mm 0mm 220mm 260mm, angle=-90, clip, scale=0.8]{hayakawa_fig6.eps}}
{\includegraphics[viewport=0mm 0mm 220mm 220mm, clip, scale=0.9]{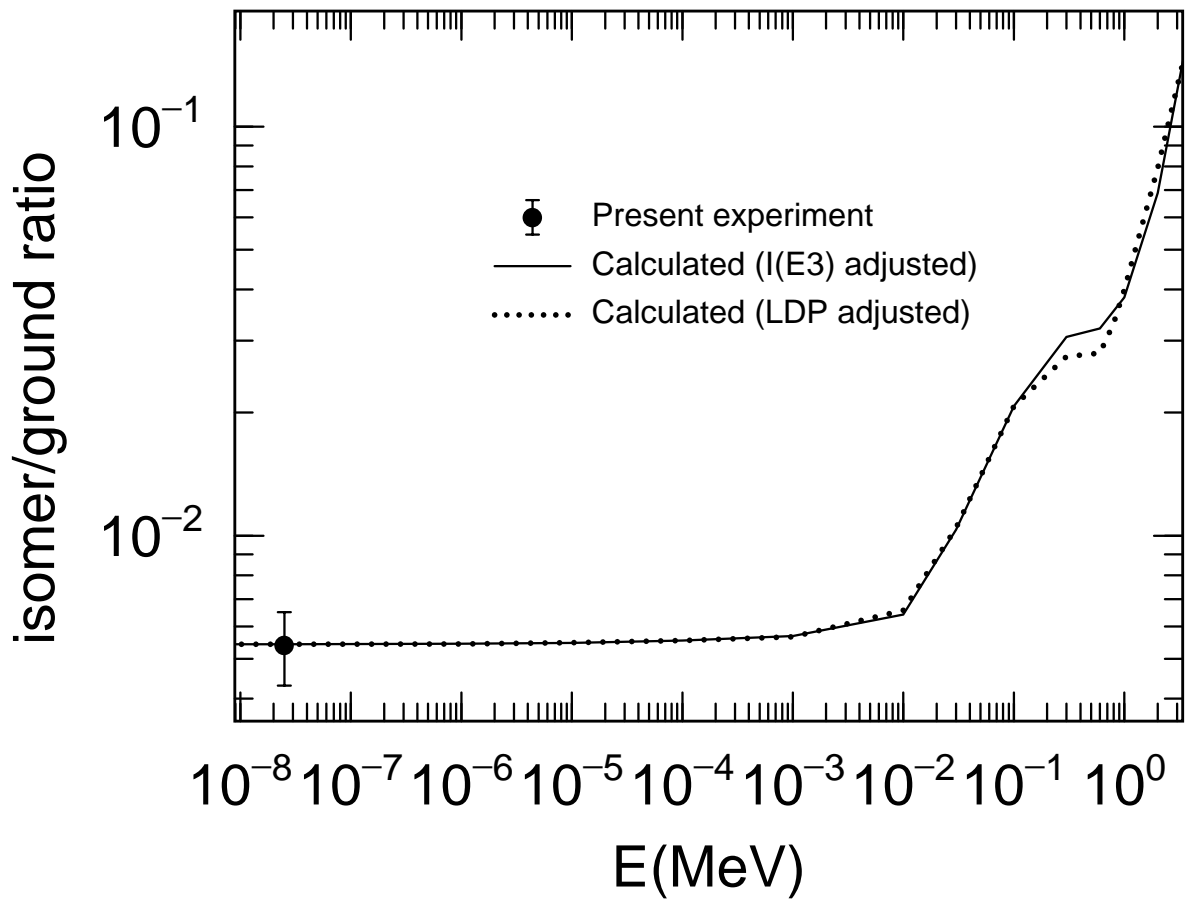}}
\caption{Cross section ratios of the isomer to the ground state of $^{186}$Re
calculated by the Hauser-Feshbach statistical-model. 
These ratios are adjusted 
to reproduce the presently measured value at the thermal neutron energy (see text).
}
\label{fig:ratio}
\end{figure}


\begin{thebibliography}{}
\bibitem[Amari et al.(1992)]{Amari92}Amari, S., Hoppe, P., Zinner, E. \& Lewis, R.S. 1992, ApJ, 394, L43
\bibitem[Aoki et al.(2003a)]{Aoki03a}Aoki, W., Honda, S., Beers, T. C. \& Sneden, C. 2003, ApJ, 586, 506
\bibitem[Aoki et al.(2003b)]{Aoki03b}Aoki, W., et al.  2003, ApJ, 592, L67
\bibitem[Arlandini et al.(1999)]{Ar99}Arlandini, C. et. al.  1999, ApJ. 525, 886
\bibitem[Arnould, \& Goriely(2003)]{Around03}Arnould, M. \& Goriely, S. 2003, PhR.., 384, 1
\bibitem[Arnould, Takahashi \& Yokoi(1984)]{Arnould84}Arnould, M., Takahashi, K. \& Yokoi, K. 1984, A\&A, 137, 51
\bibitem[Bao \& K{"a}ppeler(1987)]{Bao87}Bao, Z. Y. \& K{"a}ppeler, F.  1987, ADNDT, 36, 411
\bibitem[Baglin(2003)]{A186}Baglin, C.M. 2003, Nucl.Data Sheets {\bf 99}, 1
\bibitem[BBFH(1957)]{BBFH}Burbidge, E. M., Burbidge, G. R., Fowler, W. A., \& Hoyle, F. 1957,  ReMP., 29, 548
\bibitem[Bear(1981)]{Beer81}Beer, H., K{\"a}ppeler, F., Wisshak, K., \& Ward, R. A. 1981, ApJS, 46, 295
\bibitem[Belic et al.(1999)]{Belic99}Belic, D., et al. 1999, PhReL, 83, 5242 
\bibitem[Birck \& Allegre(1998)]{Birck98}Birck, J. L. \& Allegre, C. J. 1998, M\&PS., 33, 647
\bibitem[Bosch et al.(1996)]{Bosch96}Bosch, F.,  et al. 1996, PhReL, 77, 5190
\bibitem[Brown \& Berman(1981)]{BB81}Brown, J.C. \& Berman, B.L. 1981, PhReC 23, 1434
\bibitem[Carroll et al.(1991)]{Carroll91}Carroll, J.J., et al. 1991, PhReC, 43, 1238
\bibitem[Cayrel et al.(2001)]{Ur}Cayrel, R., et al. 2001, Nature, 409, 91
\bibitem[Clayton(1964)]{Clayton}Clayton, D.D. 1964, ApJ, 139, 637
\bibitem[Clayton(1969)]{Clayton69}Clyton, D. D. 1969, Nature, 224, 56
\bibitem[Conser \& Truran(1981)]{Cosner81}Conser, K. M. \& Truran, J. W. 1981, Ap. Space Sci., 78, 85
\bibitem[Cowan, Thielemann \& Truran(1991)]{CTT91}Cowan, J.J., Thielemann, F.-K., \& Truran, J.W. 1991, PhRe., 208, 267
\bibitem[Cowan et al.(1999)]{Co99} Cowan,J.J., et al. 1999, ApJ, 521, 194
\bibitem[Firestone(1991)]{A187}Firestone, R.B. 1991, NDS.., 62, 159
\bibitem[Firestone et al.(1998)]{ToI}Firestone, R. B., et al. 1998, Table of Isotopes, Eighth Edition, Wiley Interscience
\bibitem[Fowler \& Hoyle(1960)]{B2}Fowler, W.A. \& Hoyle. F. 1960, Ann. Phys. 10, 280
\bibitem[Gallino et al.(1998)]{Gallino98}Gallino, R., {\it et al}. 1998, ApJ 497, 388
\bibitem[Goriely \& Arnould(2001)]{Goriely01}Goriely, S., \& Arnould, M. 2001, A\&A, 379, 1113

\bibitem[Hayakawa et al.(2004)]{Hayakawa2004}Hayakawa, T., {\it et al.} 2004, PRL 93, 161102
\bibitem[Hershberger et al.(1983)]{Hershberger1983}Hershberger, R.L., {\it et al.} 1983, PhReC 28, 2249
\bibitem[Hoffman, Woosely \& Weaver(2001)]{Ho01} Hoffman, R.D., Woosely, R.D. \&  Weaver, T.A. 2001, ApJ 549, 1085
\bibitem[Holmes et al.(1976)]{Holmes1976}Holmes, J.A., Woosley, S.E., Fowler, W.A. \& Zimmerman, B.A. 1976, ADNDT, 18, 305
\bibitem[Honda et al.(2004)]{Honda04}Honda, S., Aoki, W., Kajino, T., Ando, H., Beers, T. C. Izumiura, H., Sadakane, K., Takada-Hidai, M. 2004, 604, 474
\bibitem[JENDL(2000)]{JENDL}Nuclear Data Center, Japan Atomic Energy Research Institute: "Chart of the Nuclides 2000," http://wwwndc.tokai.jaeri.go.jp/CN00/index.html (2001.12.16).
\bibitem[K{\"a}ppeler et al.(1990)]{Ka90}K{\"a}ppeler,F.,  Gallino,R.,  Busso,M.,  Picchio,G.,  Ratteri, C.M. 1990, ApJ 354, 630
\bibitem[K{\"a}ppeler et al.(1991)]{Kappeler1991}K{\"a}ppeler, F., et al. 1991, ApJ, 366, 605
\bibitem[Lambert \& Prieto(2002)]{Lambert02}Lambert, D.L.  \& Prieto, C.A. 2002, Mon. Not. R. Astron. Soc. 335, 325
\bibitem[Lee, ElEid \& Meyer(2000)]{Th00}Lee, L.-S., ElEid, M.F. \& Meyer, B.S. 2000, ApJ 533, 998
\bibitem[Lindner(1951)]{Lindner1951}Lindner, M. 1951, Phys. Rev., 84, 240
\bibitem[Lindner et al.(1986)]{Lindner1986}Lindner, M., et al. 1986, Nature, 320, 256
\bibitem[Luck, Brich \& Allegre(1980)]{Luck80}Luck, J. M., Brich, J.L. \& Allegre, C.J. 1980, Nature, 283, 256
\bibitem[Luck \& Allegre(1983)]{Luck83}Luck, J. M. \& Allegre, C. J. 1983, Nature, 302, 130
\bibitem[McEllistrem wt al.(1989)]{McEllistrem1989}McEllistrem, M.T., et al., 1989, PhReC 40, 591
\bibitem[Meyer et al.(1992)]{Meyer92}Meyer, B.S., Mathews, G.J., Howard, W.M., Woosley, S.E. \&  Hoffman, R.D., 1992, ApJ  399, 656
\bibitem[Mohr et al.(2004)]{Mohr04}Mohr, P. et al. 2004, PhReC, 69, 2801 
\bibitem[Mughabghab(1984)]{bnl325b}Mughabghab, S.F., 1984, "Neutron Cross Sections, Vol.1, Neutron Resonance Parameters and Thermal Cross Sections", Academic Press Inc.
\bibitem[Nakajima(1991)]{Nakajima91}Nakajima, Y., 1991,JNDC WG on Activation Cross Section Data: "JENDL Activation Cross Section File," Proc. the 1990 Symposium on Nuclear Data, JAERI-M 91-032, p. 43
\bibitem[Nazarewicz(1990)]{shape90}Nazarewicz, W., Riley, M. A. \& Garrett, J. D. NuPhA, 512, 61
\bibitem[Otsuki, Tagoshi \& Kajino(2000)]{Ot00} Otsuki,K.,  Tagoshi,H.,  Kajino, T., \& Wanajo, S.Y. 2000, ApJ 533, 424
\bibitem[Otsuki, Mathews \& Kajino(2003)]{Otsuki03}Otsuki, K., Mathews, G. J., Kajino, T. 2003, New Astronomy, 8, 767
\bibitem[Pellin et al.(2000)]{Pellin00} Pellin, M.J. et al. 2000, Lunar and Planet. Sci. 31, 1917
\bibitem[Prantzos et al.(1990)]{Prantzos90} Prantzos, N., Hashimoto, M. \& Nomoto, K. 1990, ApJ 234, 211
\bibitem[Qian, Vogel \& Wasserburg(1998)]{Qi98} Qian,Y.-Z., Vogel,P., \& Wasserburg, G.J. 1998, ApJ 494, 285
\bibitem[Richter, Ott \& Begemann(1998)]{Richter98}Richter, S., Ott, U. \&  Begemann, F. 1998, Nature, 391, 261
\bibitem[Rutherford(1929)]{Rutherford}Rutherford, E. 1929, Nature, 123, 313
\bibitem[Ryan, Norris \& Beers(1996)]{Ryan1996}Ryan, S.G., Norris, J.E. \& Beers, T.C. 1996, ApJ, 471, 254
\bibitem[Schaty et al.(2002)]{Schaty02}Schaty, H., {\it et al.} 2002, ApJ, 579, 626
\bibitem[Seegmiller et al.(1972)]{Seegmiller}Seegmiller, D.~W.~, et al. 1972,  NuPhA, 185, 94
\bibitem[Shizuma et al.(2003)]{Shizuma03}Shizuma, T., et al. 2003, Eur. Phys. J. A, 17, 159
\bibitem[Sneden et al.(2002)]{Sneden02}Sneden, C. et al. 2002, ApJ 566, L25
\bibitem[Sneden et al.(1996)]{Sneden96}Sneden, C., McWilliam, A., Preston, G. W., Cowan, J. J., Burris, D. L., Armosky, B. J. 1996, ApJ. 467, 819S
\bibitem[Sonnabend et al.(2003)]{Sonnabend}Sonnabend, K. et al. 2003, ApJ, 583, 506
\bibitem[Stolovt, Namenson \& Berman(1976)]{SNB76}Stolovt, A., Namenson, A.I. \& Berman, B.L. 1976, PhReC 14, 965
\bibitem[Straniero et al.(1995)]{Straniero95}Straniero, O. {\it et al.} 1995, ApJ 440, L85
\bibitem[Takahashi \& Yokoi(1983)]{Takahashi83}Takahashi, K. \& Yokoi, K. 1983, NuPhA, 404, 578
\bibitem[Utsunomiya et al.(2003)]{Utsunomiya03}Utsunomiya, H., et al. 2003, PhRvC 67, 015807
\bibitem[Wanajo et al.(2002)]{Wanajo02}Wanajo, S., et al. 2002, ApJ, 577, 853
\bibitem[Ward, Newman \& Clayton(1976)]{Ward76}Ward, R.A., Newman, M.J. \& Clayton, D.D. 1976, ApJ, 31, 33
\bibitem[Winters(1986)]{WCHH86}Winters, R.R., Carlron, R.F.., Harvery, J.A. \& Hill, N.W. 1986, PhReC, 34, 840
\bibitem[Woosley \& Fowler(1979)]{Woosley1979}Woosley, S.E. \& Fowler, W.A. 1979, ApJ 233, 411
\bibitem[Woosley \& Weaver(1995)]{Woosley95} Woosley, S. E. and Weaver, T. A. 1995,  ApJ {\bf 101}, 181
\bibitem[Woosely et al.(1994)]{Woosley94}Woosley, S.E., Wilson, J.R., Mathews,G.J.,   Hoffman, R.D., \&  Meyer, B.S., 1994, ApJ. 433, 229
\bibitem[Yin, Jacobsen \& Yamashita(2002)]{Yin02}Yin, Q., Jacobsen, S.B. \& Yamashita, K. 2002, Nature 415, 881
\bibitem[Yokoi, Takahashi \& Arnould(1983)]{Yokoi83}Yokoi, K., Takahashi, K. \& Arnould, M. 1983, A\&A, 117, 65
\bibitem[Yokoi \& Takahashi (1983)]{Yokoi83b}Yokoi, K. \& Takahashi, K. 1983, Nature,  305, 198



\end{thebibliography}
\end{document}